\documentclass[aps,prd,twocolumn,showpacs,superscriptaddress,groupedaddress,
amsmath,amssymb,floatfix,nofootinbib]
{revtex4}  
\usepackage{graphicx}
\usepackage{dcolumn}
\usepackage{bm}
\usepackage[usenames,dvipsnames]{color}
\usepackage{amsmath,amssymb}
\usepackage{slashed}
\usepackage{hyperref}

\begin{document}

%

\let\a=\alpha      \let\b=\beta       \let\c=\chi        \let\d=\delta
\let\e=\varepsilon \let\f=\varphi     \let\g=\gamma      \let\h=\eta
\let\k=\kappa      \let\l=\lambda     \let\m=\mu
\let\o=\omega      \let\r=\varrho     \let\s=\sigma
\let\t=\tau        \let\th=\vartheta  \let\y=\upsilon    \let\x=\xi
\let\z=\zeta       \let\io=\iota      \let\vp=\varpi     \let\ro=\rho
\let\ph=\phi       \let\ep=\epsilon   \let\te=\theta
\let\n=\nu
\let\D=\Delta   \let\F=\Phi    \let\G=\Gamma  \let\L=\Lambda
\let\O=\Omega   \let\P=\Pi     \let\Ps=\Psi   \let\Si=\Sigma
\let\Th=\Theta  \let\X=\Xi     \let\Y=\Upsilon
%


\def\cA{{\cal A}}                \def\cB{{\cal B}}
\def\cC{{\cal C}}                \def\cD{{\cal D}}
\def\cE{{\cal E}}                \def\cF{{\cal F}}
\def\cG{{\cal G}}                \def\cH{{\cal H}}
\def\cI{{\cal I}}                \def\cJ{{\cal J}}
\def\cK{{\cal K}}                \def\cL{{\cal L}}
\def\cM{{\cal M}}                \def\cN{{\cal N}}
\def\cO{{\cal O}}                \def\cP{{\cal P}}
\def\cQ{{\cal Q}}                \def\cR{{\cal R}}
\def\cS{{\cal S}}                \def\cT{{\cal T}}
\def\cU{{\cal U}}                \def\cV{{\cal V}}
\def\cW{{\cal W}}                \def\cX{{\cal X}}
\def\cY{{\cal Y}}                \def\cZ{{\cal Z}}

%

\newcommand{\Ns}{N\hspace{-4.7mm}\not\hspace{2.7mm}}
\newcommand{\qs}{q\hspace{-3.7mm}\not\hspace{3.4mm}}
\newcommand{\ps}{p\hspace{-3.3mm}\not\hspace{1.2mm}}
\newcommand{\ks}{k\hspace{-3.3mm}\not\hspace{1.2mm}}
\newcommand{\des}{\partial\hspace{-4.mm}\not\hspace{2.5mm}}
\newcommand{\desco}{D\hspace{-4mm}\not\hspace{2mm}}


\def\be{\begin{equation}}
\def\ee{\end{equation}}
\def\bea{\begin{eqnarray}}
\def\eea{\end{eqnarray}}
{\newcommand{\lsim}{\mbox{\raisebox{-.6ex}{~$\stackrel{<}{\sim}$~}}}
{\newcommand{\gsim}{\mbox{\raisebox{-.6ex}{~$\stackrel{>}{\sim}$~}}}
\def\mpl{M_{\rm {Pl}}}
\def\gev{{\rm \,Ge\kern-0.125em V}}
\def\tev{{\rm \,Te\kern-0.125em V}}
\def\mev{{\rm \,Me\kern-0.125em V}}
\def\ev{\,{\rm eV}}

\def\Treh{T_{R}}
\def\Tf{T_{\rm {f}}}
\def\treh{t_{\rm {reh}}}
\def\Hreh{H_{\rm {reh}}}
\def\Hmax{H_{\rm {max}}}
\def\Rmax{R_{\rm {max}}}
\def\nG{n_{\tilde G}}
\def\YG{Y_{\tilde G}}
\def\mG{m_{\tilde G}}
\def\rhoG{\rho_{\tilde G}}
\def\greh{g_{*\rm{reh}}}
\def\Sigmatot{\Sigma_{{\rm {tot}}}}

\newcommand{\vev}[1]{\langle #1 \rangle}

\newcommand{\td}{t_d}
\newcommand{\tkin}{t_{kin}}
\newcommand{\tthr}{t_{thr}}
\newcommand{\tf}{t_f}
\newcommand{\Gt}{{\widetilde{G}}}
\newcommand{\gt}{{\tilde{g}}}
\newcommand{\MeV}{\,\textrm{MeV}\,}
\newcommand{\GeV}{\,\textrm{GeV}\,}
\newcommand{\OR}{\,\textrm{or},\qquad}
\newcommand{\calA}{\mathcal{A}}
\newcommand{\half}{\frac{1}{2}}
\newcommand{\paren}[1]{\left( #1 \right)}
\newcommand{\dslash}{\slashed{\partial}}
\def\gev{{\rm \,Ge\kern-0.125em V}}
\def\tev{{\rm \,Te\kern-0.125em V}}
\newcommand{\MSUSY}{M_S}



\title{\boldmath 
Supersymmetric flat directions and resonant gravitino production}

\author{Namit Mahajan}
\email{nmahajan@prl.res.in}
\author{Raghavan Rangarajan}
\email{raghavan@prl.res.in}
\affiliation{Theoretical Physics Division, Physical Research Laboratory,
Navrangpura, Ahmedabad 380 009, India}
\author{Anjishnu Sarkar}
\email{anjishnu@lnmiit.ac.in}
\affiliation{ LNM Institute
of Information Technology, Jaipur 302031, India}
\date{\today}

\begin{abstract}
We study resonant gravitino production in the early Universe in the
presence of SUSY flat directions whose large VEVs break some but not all
gauge symmetries. We find that for a large region of parameter space the
gravitino abundance is several orders of magnitude larger than the
cosmological upper bound. Since flat directions with large VEVs are
generically expected in supersymmetric theories this result further
exacerbates the gravitino problem.
\end{abstract}

\pacs{98.80.Cq,12.60.Jv}
\maketitle

\section{Introduction}

Supersymmetric theories generically admit a large landscape of moduli
space or flat directions along which the potential vanishes classically
\cite{Affleck:1983mk}.  The flat directions can be described in terms
of gauge invariant monomials that are built out of chiral superfields
$\phi_k$ subject to specific constraints originating due to F- and D-flat
requirements; in the context of the Minimal Supersymmetric Standard
Model (MSSM), flat directions have been 
catalogued in \cite{Dine:1995kz,Gherghetta:1995dv}.
A flat direction can be represented by a modulus field $\phi$ and the
different SUSY preserving vacua along the flat direction, i.e. different
choices of the flat direction VEV, are not physically equivalent.
Supersymmetry breaking lifts the flat directions, and in the early
Universe $\phi$ can be displaced away from the origin with a large vacuum
expectation value (VEV).

Non-zero VEVs for flat directions typically break one or more gauge
symmetries, and the corresponding gauge supermultiplet acquires
a mass $\propto \varphi$, where $\varphi$ is the VEV of $\phi$.
Since $\varphi$ can be very large, scattering processes mediated by
the heavy gauge bosons get suppressed, and the thermal history of
the Universe can be very different from the standard thermal history
of the Universe (see Ref. \cite{Enqvist:2003gh}, and references therein).  
For example, if 
a flat direction associated with a squark field gets a VEV it breaks all
gauge symmetries.  This then leads to a delay in thermalization after
inflation 
which can suppress gravitino production
\cite{Allahverdi:2005fq,Allahverdi:2005mz,Rangarajan:2012wy}.

{
If, on the other hand, the flat direction under consideration preserves
some of the gauge symmetries, then there can be reasonably fast
thermalisation.  In this article we suggest a new mechanism for 
enhanced gravitino
production 
in the presence of a SUSY
flat direction with a large VEV
in the context of a thermal Universe.  We find that in a large 
region of the parameter space the gravitino abundance is several orders of
magnitude larger than the cosmological upper bound. 
For concreteness we
consider the specific flat direction $H_uH_d$. This breaks $SU(3)_C
\times SU(2)_L
\times U(1)_Y \rightarrow SU(3)_C\times U(1)_{\mathrm{EM}}$. This means that the
gluon and gluino, and photon and photino do not get mass due to large
$\varphi$ (allowing for thermalisation) while other particles coupled to the flat direction
get a contribution to their mass proportional to $\varphi$. 
As we discuss below, there can be large resonant gravitino production in such a
scenario when the intermediate particle in the s-channel process goes on the mass shell.
}

{ At finite temperature, the SUSY breaking scale,
$\MSUSY$,
and the mass splitting between a particle and its superpartner,
are set by the temperature $T$ of the thermal bath (see for example \cite{Girardello:1980vv}). 
Then $M^2-m^2=\delta\, T^2+m_0^2$, where $\delta$
denotes the splitting due to the finite temperature between the square of
the sfermion mass $M$ and the fermion mass $m$ in units of $T^2$, and
$m_0$ is the zero temperature soft SUSY breaking parameter.  In our
scenario, the thermal splitting dominates over $m_0^2$.  
%
%

We further assume that
$\varphi \gg T $.  Quarks and charged leptons
and their superpartners get a contribution to their mass $\propto\varphi$. 
Of these particles, those with a small Yukawa 
coupling to the flat direction will still be relativistic while others will be heavy. 

$W$ and $Z$ and their superpartners will be heavy due to 
their coupling with the flat direction.
The photino and gluino get a  mass $\propto T$ due to SUSY breaking as above.   
The gravitino is
much lighter.  
{The gravitino mass
is $m_{\tilde G}
\sim \MSUSY^2/M_{P} =(\delta' T^2/M_P)$ where $M_P=M_{Pl}/ \sqrt{8\pi}\simeq
2.4\times 10^{18}$ GeV is the reduced Planck mass, and we 
take $\delta'\sim0.1$.} 

Now consider the
following scattering reaction: $\tilde{A} + f \longrightarrow \tilde{f}^*
\longrightarrow f + \tilde{G},$ where $f\,, \tilde{A}\,, \tilde{G}\,,
\tilde{f}$ denote a charged (heavy) fermion, gluino/photino, gravitino and
sfermion respectively. (For example, one could consider quark and gluino scattering to
quark and gravitino.)  
Since 
$\varphi \gg T \gg m_0$ implying
$\delta
T^2, m_0^2\ll m^2$, the initial state fermion and intermediate sfermion are
almost degenerate in mass.  Moreover, the sfermion in the s-channel exchange
can be on the mass shell.  This Breit-Wigner resonance then gives a 
large contribution
to the scattering cross section and a very large abundance of gravitinos.
The Boltzmann suppression of the incoming heavy fermion is compensated by
the Breit-Wigner resonance factor.

Neglecting other contributions, the s-channel resonant production cross section 
is given by (considering only the helcity $1/2$ component for the gravitino)
\bea
\sigma(s)
&\approx&
\frac{1}{3.8.2.2}\frac{2N_g}{4}
\frac{(M^2-m^2)^2}{3\mG^2 M_P^2}
\frac{\alpha_g}{s}
\frac{(s-m^2)^2}{(s-M^2)^2+M^2\Gamma^2}
%
\eea
where $\Gamma\ll M$ is the width of the intermediate on-shell sfermion,
$\alpha_g = g^2/(4\pi)$ where $g$ is the relevant gauge coupling, and
we have used $\mG\ll m,M$.  
Feynman rules for gravitino
interacations are in Ref. \cite{Moroi:1995fs},
and $N_g$ is $\sum_A Tr[T^A T^A]$ where $T^A$
is the generator of the relevant gauge group.  
Since we are considering the goldstino part of the
gravitino, which comes from SUSY breaking, 
the cross section should not be $M_P$ suppressed (see for example \cite{Martin:1997ns}).
%
Using $m_\Gt M_P= \delta'  T^2$
and
$(M^2-m^2)=\delta T^2$ and assuming
$\delta\sim\delta'$ we get
\be
\sigma(s)
\approx
\frac{N_g}{576}
\frac{\alpha_g}{s}
\frac{(s-m^2)^2}{(s-M^2)^2+M^2\Gamma^2}
\ee
We let $\Gamma=M/z$ and take $z=50,500$ in our analysis
(though representative values for our scenario below may be 
much smaller).
Lower values of $\Gamma$ can increase $\sigma$.

This should be contrasted with the zero temperature case, say within gravity mediated
scenario where $M^2-m^2 \sim m_0^2 \ll m_\Gt M_P =\MSUSY^2$ leading to a strong suppression
factor. We emphasize again that this new and novel feature is due to distinctive character
of supersymmetric theories at finite temperature.} 

The mechanism considered above is different from enhanced gravitino production
during preheating which has been considered in
Ref.~\cite{giudiceetal1,maroto.00,kallosh.00,tsujikawa.00,
nilles.01,Nilles:2001fg,
nilles&olive.01,Greene:2002ku,podolsky}.

\section{Boltzmann equation}

Gravitinos are produced by the scattering of the decay products of the
inflaton 
\cite{Nanopoulos:1983up,Krauss:1983ik,Falomkin:1984eu,Khlopov:1984pf,
Ellis:1984eq,Juszkiewicz:1985gg,Ellis:1984er,Kawasaki:1986my,
Khlopov:1993ye,Moroi:1993mb,Kawasaki:1994af,Bolz:2000fu,
Cyburt:2002uv,Giudice:1999am,Kawasaki:2004qu,Pradler:2006qh,
Pradler:2006hh,Rangarajan:2006xg,Rangarajan:2008zb,Rychkov:2007uq}.
Refs. \cite{Ellis:1984eq,Kawasaki:1994af}
provide a list of processes for gravitino production in the standard
scenario for thermal gravitino production.

The number density of a species $X_3$ participating in reactions
$X_1 X_2 \rightleftharpoons X_3 X_4$ can be obtained via
the integrated Boltzmann equation,
\be
\dot{n}_3 + 3 H n_3 = {\cal C}
\ee
where $\cal C$ is the collision integral.
When the number density of $X_3$ is small, as we presume in our case where $X_3$
represents the gravitino, 
we can ignore the $X_3 X_4\rightarrow X_1 X_2$ process.  Then
\begin{align}
\dot{n}_3 + 3 H n_3 = 
\int ~d\Pi_1 ~d\Pi_2 
~f_1 ~f_2 ~W_{12}(s) 
\equiv A,
\label{eq:mbltzeq} 
\end{align}
where $f_i$ are phase space distribution functions and 
$d\Pi_i \equiv \frac{g_i}{(2\pi)^3} \frac{d^3p_i}{2E_i}$.
%
%
$g_i$ is the number of internal degrees of freedom of species $i$.
Then, 
from Ref. \cite{Edsjo:1997bg},
\be
A = \frac{T}{32\pi^4}\sum_{1,2}\int \,ds\, g_1 g_2 p_{12} W_{12} 
K_1\left(\frac{\sqrt s}{T}\right)\,, \label{eqA}
\ee
where $W_{12}(s) =
4 p_{12} ~\sqrt{s} ~\sigma_{CM}(s)$,  
%
$\sigma_{CM}$ is the cross section in the centre-of-mass frame and
\begin{equation}
p_{12} = \frac{\left[ s - (m_1+m_2)^2 \right]^{1/2} 
\left[ s- (m_1-m_2)^2 \right]^{1/2}}{2\sqrt{s}} 
\label{eq:p12} 
\end{equation}
is the magnitude of the momentum of particle $X_1$ (or $X_2$) in the
center-of-mass frame of the particle pair $(X_1,X_2)$. $K_1$ is the
modified Bessel function of the second kind of order 1.  
Note that its
exponential decay at large $s$
provides the Boltzmann suppression associated with the incoming heavy quark.
\footnote{
{The
derivation of $A$ presumes a Maxwell-Boltzmann distribution for both
incoming particles, while our gluino is relativistic.  However it
has been argued in Ref. \cite{Scherrer:1985zt} that final abundances
are insensitive to the statistics.\\} 
We ignore gravitino decay
in the Boltzmann equation as the gravitino lifetime is $~10^{7-8}
(100\GeV/m_\Gt)$\,s \cite{Ellis:1984eq} and is not relevant during the
gravitino production era.} 
For $p_{12}$, we get $(s-m^2)/(2\sqrt {s})$ for the incoming gaugino 
mass smaller than $T$ and hence much smaller than $m$.

Substituting $\sigma_{CM}$ in $A$, we need to do the integral over $s$.
We shall specifically consider the process 
\be
\tilde{g} + q \longrightarrow \tilde{q}^* \longrightarrow q + \tilde{G}\,.
\ee
For this process $g_1=2\times8$ and $g_2=2\times3$ and $\alpha_g$
is replaced by $\alpha_s$.  Throughout we shall ignore the variation
of $\alpha_s$ with temperature and take $\alpha_s=5\times10^{-2}$,
as relevant for the temperatures in our scenario.

\section{Resonant gravitino production}

For obtaining $A$ we first
discuss the evolution of $M\sim
m = h \varphi$, where $h$ is a relevant Yukawa coupling.  \footnote{When
$\phi$ is oscillating, $\varphi$ is the amplitude of oscillation, as the
period of the oscillation $m_\phi^{-1}$ is much smaller than the timescale
for gravitino production.} We take the mass of the flat direction $m_\phi$
to be related to the scale of SUSY breaking. 
{
Immediately after inflation, when the Universe is cold, $m_\phi=m_0$.
When 
$H$ decreases to
$H\sim m_\phi=m_0$ at $t_0\sim1/m_0$, $\phi$ starts oscillating
and thereafter $\varphi$ decreases as $1/a^{3/2}$.  
($a$ is the scale factor of the Universe.)
Subsequently at $t_d$ the inflaton decays, the temperature becomes $T_R$ (in
the instantaneous decay approximation) and then the temperature also determines
the SUSY breaking scale and the mass of the 
flat direction:
} 
$m_\phi^2=h^{\prime 2} T^2+m_0^2$, where $h'$
is the Yukawa coupling for some light field in thermal equilibrium.
\footnote{Thermal corrections to the flat direction potential of the form
$h^{\prime 2}T^2|\phi|^2$ and $\alpha_g^2 T^4 \log(|\phi|^2)$, along with
non-renormalisable terms, have been considered 
in Refs. \cite{Allahverdi:2000zd,Anisimov:2000wx,Anisimov:2001dp}.
We presume 
that thermal corrections to the flat direction potential is effectively
quadratic with a contribution of $h^\prime T$
to the mass.
For the light field with Yukawa coupling $h^\prime$ to be
relativistic and in thermal equilibrium, 
its mass $h^\prime\varphi(t)$ should be less than $T(t)$.} 
$H=10 T^2/M_{Pl}<m_\phi$ and $\phi$  continues to oscillate after $t_d$.
The oscillating field can be thought of as a condensate of zero momentum
particles.

We take the initial VEV of $\phi$ at $t_0$ to be $\varphi_0$. 
Then for  $t >
\td = \Gamma_d^{-1}$, where $\Gamma_d$ is the inflaton decay rate,
the quark mass is given by
\begin{equation}
m^2 = h^2\varphi_0^2 
\paren{\frac{a_0}{a_d}}^3 \paren{\frac{a_d}{a}}^3
= 
h^2\varphi_0^2 \paren{\frac{\Gamma_d}{m_0}}^2 \paren{\frac{T}{T_R}}^3
\label{squarkmass1}
\end{equation}
where we have used $a \sim t^{2/3}$ for $t_0 < t < t_d$ for an inflaton
oscillating in a quadratic potential during reheating and $a \sim 1/T$ for
$t > t_d$. $T_R$ is the reheat temperature at $t_d$ and
is given by \cite{Kolb:1990vq}
\be
T_R=0.55 g_{**}^{-1/4}\Gamma_d^{1/2}\mpl^{1/2}
\ee
where $g_{**}$ is the number of relativistic degrees of freedom relevant when
the flat direction VEV is large and many species are non-relativistic.  Taking
the relativistic species to be
the photino, photon, gluino and gluon,
$g_{**}=33.75$. We further define $m_d\equiv m(t_d)=m_{t0}(\Gamma_d/m_0)$, where
$m_{t0}=h\varphi_0$.

{
After the inflaton decays, the energy density $\rho_\phi$ in the flat
direction condensate is $\half m_\phi^2 \varphi^2$ while the energy
density of the radiation $\rho_{rad}=(\pi^2/30)g_{**}T^4$.  
For the parameter values we consider below, the Universe is radiation dominated after inflaton decay,
and therefore}
\begin{equation}
T=T_R\left(\frac{t_d}{t}\right)^\half
\end{equation}
$A$ in Eq. (\ref{eqA}) is a function of $T$.  
$A$ can now be expressed as a function of $t$ and we can solve 
Eq. (\ref{eq:mbltzeq}) to obtain the number density of gravitinos.
We will finally like to obtain the gravitino number density
at $t_{e}$ when the flat direction condensate decays, or the resonant
mechanism terminates.

The condensate decays (perturbatively) at $t_f$ when its decay rate 
$\Gamma_\phi=m_\phi^3/\varphi^2$ equals $H$ 
\cite{Affleck:1984fy,Olive:2006uw}.
(We discuss alternate mechanisms for condensate decay below.)
Then at any temperature, for $m_\phi\sim h'T$
\be
\Gamma_\phi=h^{\prime 3}T^3/\varphi^2=h^{\prime 3}T^3/[\varphi_d^2 T^3/T_R^3]=
h^{\prime 3}T_R^3/\varphi_d^2
\ee
using $\varphi^2\sim 1/a^3\sim T^3$.  Now $\varphi_d^2=\varphi_0^2
(a_0/a_d)^3=\varphi_0^2 (t_0/t_d)^2=\varphi_0^2(\Gamma_d/m_0)^2.  $ Then,
\be t_f=\frac{\varphi_0^2 \Gamma_d^2}{h^{\prime 3}T_R^3 m_0^2}\,.  \ee

Let $T_m$ be the temperature when the condensate thermal mass $h'T$ equals
$m_0$.  If $t_f$ obtained above is greater than $t_m=t_d(T_R/T_m)^2$,
then one should use $m_\phi=m_0$ to obtain $t_f$.  $t_f$ is then
obtained as in Eq. (46) of Ref. \cite{Rangarajan:2012wy} as 
\be t_f=
\frac{\varphi_0^{4/5}\Gamma_d^{1/5}}{m_0^2} 
\ee 
In our numerical analysis below the thermal mass $h'T$ for $\phi$
is less than $m_0$
at $t_d$ itself for the low reheat temperature that we consider.

It may happen that the resonant phenomena breaks down before $t_f$
at some time $t_r$.  We require the sfermion and fermion masses to be
much larger than $T$.  Now the quark mass $\sim T^{3/2}$ and so falls
faster than the temperature.  Defining $T_r$ via $m(T_r)=T_r$ and using
Eq. (\ref{squarkmass1}) we get
\begin{equation}
 T_r = \left(\frac{m_0}{\Gamma_d}\right)^2 \left(\frac{\Treh}{m_{t0}}\right)^2 \Treh 
 = \left(\frac{T_R}{m_d}\right)^2 \Treh\,.
\end{equation}
%
%
As $t\sim a^2\sim1/T^2$ for $t>t_d$,
\bea
t_r &=& \left(\frac{m_d}{\Treh}\right)^4 t_d
\eea

The final gravitino abundance is the abundance at $t_e={\rm min}(t_f,t_r)$
when resonant gravitino production ends and is given by 
\bea
Y(t_e)&\equiv& \frac{n(t_e)}{s(t_e)}
\label{abundance}
\eea
where $s$ is the entropy density.  We obtain the gravitino number
density by solving the integrated Boltzmann equation till $t_e$. Now the
temperature at $t_e$ just after the resonant gravitino production ends
is $T_e^\prime=(g_{**}/g_*)^{1/4} T_e$, where $T_e=T_R(t_d/t_e)^{1/2}$
is the temperature just before the end of resonant gravitino production.
Then the entropy density is $s(t_e) = ({2\pi^2}/{45})\,g_* T_e^{'3} $
We take $g_*=228.75$.  Note that the energy density in $\phi$ is sub-dominant
when the flat direction decays for the cases considered below.

After the flat direction condensate decays the gravitino mass is
given by the expression relevant to the mechanism of supersymmetry
breaking.  In gravity mediated supersymmetry breaking, $m_{3/2}\sim
m_0\sim 100-1000\gev$.  The abundance obtained above can then be
compared with the corresponding upper limit of $10^{-14}$ obtained in
Ref. \cite{Cyburt:2009pg} from various cosmological constraints for
$m_{3/2}\sim 100\gev$.  For $m_{3/2}\sim 1000\gev$ the upper limit
is $10^{-16}$.

In addition to resonant gravitino production involving heavy
quarks and squarks, gravitinos are also produced by the usual
non-resonant thermal scattering of relativistic particles
during reheating \cite{Giudice:1999am,Pradler:2006hh, Kawasaki:2004qu,
Rangarajan:2006xg,Rangarajan:2008zb,Rychkov:2007uq} and after reheating.
The total abundance generated will be proportional
to $T_R$.  We shall choose $\Gamma_d$ such that the reheat temperature
is low enough $(\le 10^6\gev)$ to suppress gravitino production via
non-resonant thermal production.

\section{Results}

We now consider plausible values of $\varphi_0$.  The non-zero vacuum
energy during inflation breaks SUSY and can give large positive
masses of order $H_I$ to the flat direction, where $H_I$ is the
Hubble parameter during inflation \cite{Dine:1995uk,Gaillard:1995az}.
Then quantum fluctuations during inflation give a VEV of order $H_I$
\cite{Dine:1995uk}.  Assuming that the field does not vary much till
$t_0$, $\varphi_0\sim H_I\le 10^{13}\gev$.

Alternatively, in some theories the contribution to the flat
direction potential during inflation due to $H_I$ is negative at the
origin \cite{Gaillard:1995az,Dine:1995uk}.  This correction to the
potential, along with non-renomalizable terms, leads to a shifted
minimum of the potential.  Then one obtains a large VEV of order
$M_{Pl}$ \cite{Gaillard:1995az}, or $10^{12-14} \GeV$ on including GUT
interactions \cite{Campbell:1998yi}.  The shifted minimum of the potential
for $\phi$ is $\Lambda(H/\Lambda)^{1/(n+1)}$ for non-renormalizable
terms of the form $\phi^{2n+4}/\Lambda^{2n}, n \ge 1$, where $\Lambda$
is the scale of some new physics \cite{Campbell:1998yi,Dine:2003ax}.
$\phi$ oscillates about this time dependent minimum which decreases
as $H$ decreases.  When $H \sim m_0$ at $t_0$, the potential minimum
goes to zero and the field oscillates about the origin in a quadratic
potential with curvature $m_0^2$.  Then $\varphi_0 \sim
\Lambda ( H(t_0) / \Lambda )^{1/(n+1)} \,,$
%
%
where $H(t_0) = m_0$  \cite{Dine:2003ax}. If $\Lambda\sim 10^{16}\gev$, or $M_{\rm Pl}$, then
we get $\varphi_0\sim10^9\gev$, or $3\times 10^{10} \gev$, for $n=1$.
For larger $n$, $\varphi_0$ will be larger.

For our analysis below we present a few cases with different parameter
values. We take $\Gamma_d=10^{-6}\gev$ to ensure that the reheat
temperature of $8\times10^5\gev$ is below the upper bound of $10^6\gev$
for (non-resonant) thermal gravitino production \cite{Kawasaki:2008qe}.
$t_d$ is then $10^6\gev^{-1}$.

For $\varphi_0=10^{16}\gev$ we take $z=50$, $\delta=0.1$, $h=0.5$,
$h^\prime=10^{-5}$ and $m_0=100\gev$.  For $\varphi_0=10^{17}\gev$
we take $z=500$, $\delta=0.1$, $h^\prime=10^{-5}$, $m_0=100\gev$ with
$h=0.1$ and $m_0=1000\gev$ with  $h=0.2$.  For the first case we obtain
$Y=1\times10^{-5}$ at $t_{e}=4\times 10^7\gev^{-1}$.  For the second case
we obtain $Y=3\times10^{-8}$ at $t_{e}=3\times 10^8\gev^{-1}$.  For the
third case we obtain $Y=7\times10^{-4}$ at $t_{e}=3\times 10^6\gev^{-1}$.
All these abundances are much larger than the cosmological upper bounds
of $10^{-14,-16}$ mentioned above.

We further point out that for $\varphi_0\le10^{15}\gev$ the collision
integral on the r.h.s. of the integrated Boltzmann equation is so
large that one gets an abundance much larger than 1.  While this
is in conflict with the assumption of a small gravitino abundance
presumed while obtaining Eq. (\ref{eq:mbltzeq}), it indicates that the
gravitino number density would be equal to the equilibrium gravitino
number density in such cases.  The abundance at $t_{e}$ is then $
Y(t_{e})= {n_\Gt^{eq}(t_{e})}/{s(t_{e})} \approx 8\times10^{-3}
$, where the equilibrium gravitino number density $n_\Gt^{eq}(t_e)=
3 \,\zeta(3)/(4\pi^2)\, 2\, T_{e}^3$.

The gravitino abundance is larger for smaller $m_d$ and larger $\delta$
as they increase the phase space available for resonant production.
Because $m$ and $M$ are very close in mass, $M-m\approx(\delta
T^2+m_0^2)/(2m)\ll \Gamma/2$, the initial value of $\sqrt s$ lies within
the Breit-Wigner peak in the integral over $s$ in the cross section.
Increasing $\delta$, or decreasing $m_d$, allows one to sample more of the
Breit-Wigner resonance and thus gives a larger contribution.  $m_d$ is a
function of $h$, $\varphi_0$, $\Gamma_d$, and $m_0$.  
Decreasing $h$, $\varphi_0$ or $\Gamma_d$,
or increasing $m_0$ (which makes the condensate oscillate earlier),
decreases $m_d$ and increases the gravitino abundance.  At later times
one samples more of the Breit-Wigner resonance as $\delta T^2\sim 1/a^2$
while $m\Gamma\sim 1/a^3$.

Even though $A$ contains a Boltzmann suppression factor because of the heavy
incoming quark, the resonance effect overcomes this suppression, as mentioned
earlier.  
We have verified that for incoming energies away from resonance the gravitino
production cross section is indeed suppressed.

We have only considered one channel for gravitino production for this
flat direction.  One can consider processes involving other particles such
as photinos and charged leptons.  Other flat directions with large VEVs
can also lead to resonant gravitino production.  For example, the flat
direction parametrised by the monomial $LLe$ will break $SU(2)_L\times
U(1)_Y$.  Gluons and gluinos could then participate in resonant gravitino
production as above.

\section{Discussion and Conclusion}

Our results indicate that there can be large gravitino production through
a resonant process in a thermal Universe in the presence of a large VEV
for a SUSY flat direction that breaks some but not all gauge symmetries.
For the parameters considered in the previous section we find that the
gravitino abundance exceeds the cosmological upper bounds, and in many
cases can equal the large equilibrium abundance.  Since large VEVs for
SUSY flat directions is a generic feature in supersymmetric cosmological
scenarios our results are very relevant to the understanding of the
gravitino problem in the early Universe.

Lowering the reheat temperature (by decreasing $\Gamma_d$)
increases the gravitino abundance.
This is in contrast with the standard non-resonant thermal production
scenario in which the abundance is proportional to the reheat temperature.
This implies that if we consider lowering the reheat temperature, the
standard solution to the gravitino problem, it will lead to even more
gravitino production from the resonant scattering process discussed above.

One mechanism to
decrease the large gravitino abundance obtained above is to invoke the quick
decay of the flat direction.  The longevity of flat directions has been
debated in Refs. \cite{Allahverdi:1999je,Postma:2003gc,Olive:2006uw,
Allahverdi:2006xh,Basboll:2007vt,Basboll:2008gc,Allahverdi:2008pf,
Gumrukcuoglu:2008fk,Gumrukcuoglu:2009fj,Allahverdi:2010xz}.  However it
has been argued in Refs. \cite{Allahverdi:2008pf,Allahverdi:2010xz}
that even if non-perturbative rapid decay via parametric resonance occurs
for scenarios with multiple flat directions it leads to a redistribution
of energy of the condensate amongst the fields in the D flat superspace
and hence to practically the same cosmological consequences, including
at least as large masses as in the scenario with only perturbative decay.

Scattering of particles of the thermal bath off the flat direction
condensate can lead to the decay of the condensate \cite{Dine:1995kz,
Allahverdi:2000zd,Anisimov:2000wx}, though thermal effects are less
important for larger values of $n$. For example, for $n=3$ the condensate
decays much after the decay of the inflaton \cite{Anisimov:2000wx}. Decay
via fragmentation into solitonic states called Q-balls
\cite{Kusenko:1997si,Enqvist:1997si,Kasuya:1999wu,Enqvist:2000gq,
Kasuya:2000sc,Kasuya:2000wx,Kasuya:2001hg,Enqvist:2000cq,Multamaki:2002hv}
or Q-axitons \cite{Enqvist:1999mv} due to inhomogeneities in the
condensate may also be relevant. The relevant time scale for Q-ball
and Q-axiton formation is $10^{2-4} m^{-1}$, where $m$ is a mass scale
associated with the flat 
direction \cite{Enqvist:1997si,Kasuya:2000wx,Kusenko:1997si,Enqvist:1999mv}, 
which can decrease the lifetime of
the flat directions considerably to even less than $t_d$.
However, Q-balls/axitons may not form if there is no related conserved charge
associated with the flat direction (usually baryon or lepton number).

In conclusion, we have pointed out that there could be excessive
gravitino production in the early Universe through a resonant mechanism
in the presence of flat directions in supersymmetric theories.  The final
abundance can exceed the cosmological bound on the gravitino abundance by
several orders of magnitude.  This result would be relevant for typical
supersymmetric scenarios of the early Universe, and exacerbates the
well known gravitino problem.  Mechanisms for the quick decay of the
flat directions may need to be invoked to suppress the final gravitino
abundance.

\end{document}